\documentclass[12pt]{article}

\usepackage{color}
\definecolor{darkblue}{rgb}{0.1,0.1,.7}
\usepackage[colorlinks, linkcolor=darkblue, citecolor=darkblue, urlcolor=darkblue, linktocpage]{hyperref}
\usepackage[]{amsmath}
\usepackage[]{graphicx}
\usepackage[]{latexsym}
\usepackage{slashed,graphicx,color,amsmath,amssymb}
\usepackage[mathscr]{eucal}
\usepackage{mathrsfs}
\usepackage{geometry}
\usepackage[margin=10pt,font=small,labelfont=bf]{caption}
\usepackage{amscd}
\usepackage{bm}
\usepackage{xcolor}
\usepackage{upgreek}
\usepackage{caption}
\usepackage{subcaption}
\usepackage[square, comma, sort&compress,numbers]{natbib}
\usepackage[all,cmtip]{xy}
\numberwithin{equation}{section}
\newcommand{\tr}{\mathrm{Tr}\,}

\newcommand{\rP}{\mathrm{P}}

\begin{document}
\vspace*{-.6in} \thispagestyle{empty}
%\begin{flushright}
%\textcolor{gray}{Personal Notes}
%\end{flushright}
\vspace{.2in} {\Large
\begin{center}
{\bf $T\bar{T}$-deformation: a lattice approach}
\end{center}}
\vspace{.2in}
\begin{center}
Yunfeng Jiang
\\
\vspace{.3in}
\small{
\textit{
School of Phyiscs and Shing-Tung Yau Center\\ Southeast University, Nanjing 210096, China
}
}

\end{center}

\vspace{.3in}

\begin{abstract}
\normalsize{Integrable quantum field theories can be regularized on the lattice while preserving integrability. The resulting theory on the lattice are integrable lattice models. A prototype of such a regularization is the correspondence between sine-Gordon model and 6-vertex model on a light-cone lattice. We propose an integrable deformation of the light-cone lattice model such that in the continuum limit we obtain the $T\bar{T}$-deformed sine-Gordon model. Under this deformation, the cut-off momentum becomes energy dependent while the underlying Yang-Baxter integrability is preserved. Therefore this deformation is integrable but non-local, similar to the $T\bar{T}$-deformation of quantum field theory. }
\end{abstract}

\vskip 1cm \hspace{0.7cm}

\newpage

\setcounter{page}{1}
\begingroup
\hypersetup{linkcolor=black}
\tableofcontents
\endgroup

%%%%%%%%%%%%%%%%%%%%%%%%%%%%%%%%%%%%%%%%%%%%%%%%%%%%%%%%%%%%%%%%%%%%%%%%
\section{Introduction}
\label{sec:intro}
%%%%%%%%%%%%%%%%%%%%%%%%%%%%%%%%%%%%%%%%%%%%%%%%%%%%%%%%%%%%%%%%%%%%%%%%
Understanding ultra-violate behavior of quantum field theory is a fundamental question. Typical UV behavior includes the famous Landau pole, which indicates that UV completion is needed; or asymptotic safety which means the theory flows to a UV fixed point. Recently, a third paradigm was proposed and dubbed as \emph{asymptotic fragility} \cite{Dubovsky:2012wk,Dubovsky:2017cnj}. The novel feature of this new behavior is that the UV theory is well-defined, but not local in the usual sense. Therefore it is not a fixed point like CFT, but some non-local theory.\par

Intriguingly, such non-local theories can be constructed by special irrelevant deformations of usual quantum field theories. The most studied example of such deformations is the $T\bar{T}$-deformation \cite{Smirnov:2016lqw,Cavaglia:2016oda}. It is discovered that the deformation has many special features. Most notably, it preserves integrability and is solvable in a certain sense. This allows analytic computations for many physical quantities, especially when the original theory is a CFT or integrable quantum field theory (IQFT).\par

Despite many developments, some basic and important questions remain open. One outstanding question is that the precise nature of the non-locality is not completely understood. This is reflected in the computation of correlation functions of local operators. On the one hand, in a truly gravity theory, these quantities are in general not well-defined since they are not diffeomorphism invariant. On the other hand, the $T\bar{T}$-deformed QFT is simpler than coupling the theory to generic gravity theories as the gravity theory it couples to is quite simple. It is thus an open question whether one can make sense of these quantities in the deformed theory and if so, how to compute them non-perturbatively. Very recently, there has been progress in computing correlation functions using various methods \cite{Cardy:2019qao,Cui:2023jrb,Aharony:2023dod,Castro-Alvaredo:2023rtl,Castro-Alvaredo:2023wmw}. For CFTs, the deformed correlators are most conveniently written in momentum space. Performing a Fourier transform back to spacetime is subtle \cite{Cardy:2019qao,Cui:2023jrb,Aharony:2023dod}. For IQFTs, one can adapt the form factor bootstrap approach \cite{Castro-Alvaredo:2023rtl,Castro-Alvaredo:2023wmw}. So far explicit result has been obtained up to two-particle contributions. It already exhibits some new features. In particular, for one sign of the deformation parameter the form factor expansion does not converge, even at two-particle level. \par

To proceed further requires a better understanding of the UV behavior of the deformed theories. To this end, we advocate another non-perturbative approach for studying $T\bar{T}$-deformation. A well-known approach to tame the UV divergences of a QFT is putting the theory on the lattice. The lattice regularization not only removes UV divergences, but also builds a fruitful connection between quantum field theory and statistical mechanics. This approach is particularly suitable for integrable quantum field theories because the lattice regularization can be performed in an integrability-preserving way. The resulting theory on the lattice are integrable and can be solved by Bethe ansatz. One of the most well-known example is the lattice regularization of sine-Gordon theory where the corresponding lattice model is the six-vertex model \cite{Destri:1987ze,Destri:1992qk,Destri:1994bv,Reshetikhin:1993wm}. The latter is intimately related to the Heisenberg XXZ spin chain.\par

Given a lattice regularization of IQFT, a natural and intriguing question is how to deform the underlying integrable lattice model in such a way that in the continuum limit we obtain the $T\bar{T}$-deformed quantum field theory. Such a deformation, if exist, must preserve integrability and should be non-local. Once this deformation is found, we can gain more intuition about $T\bar{T}$-deformation and understand better what happens in the UV. In addition, some physical quantities such as correlation functions can be computed on the lattice using integrability techniques. By taking continuum limit, one can compute these quantities in the corresponding field theory non-perturbatively.\par

To find the deformation for the lattice model, we investigate the light-cone lattice regularization of the sine-Gordon theory. This is a prototype of IQFTs, it is interesting both classically and in the quantum regime. At the classical level, its equation of motion leads to the famous sine-Gordon equation, which allows soliton solutions. Integrability at the classical level is guaranteed by the Lax representation and the classical Yang-Baxter equation. Soliton solutions can be found systematically by inverse scattering method. The $T\bar{T}$-deformation at the classical level has been investigated in various works, see for example \cite{Conti:2018tca,Chen:2021aid}.\par 

At the quantum level, it is known that $T\bar{T}$-deformation modifies the $S$-matrix by multiplying a simple and universal CDD (short for Castillejo-Dalitz-Dyson \cite{Castillejo:1955ed}) factor \cite{Smirnov:2016lqw,Cavaglia:2016oda}. This affects the finite-volume spectrum of the model in a simple way. The spectrum of the deformed sine-Gordon model can be calculated by the non-linear integral equation (NLIE) approach. By comparing lattice calculations with the $T\bar{T}$-deformed NLIE, we can `trace back' the deformation to the lattice model. The answer is simple. We need to deformed the cut-off rapidity of the lattice model in an energy-dependent way. In the continuum limit we keep the mass scale fixed, then the deformation of cut-off rapidity is equivalent to deforming the lattice spacing in an energy-dependent way. This echoes the fact that $T\bar{T}$-deformation amounts to deform the radius in an energy-dependent way. Away from the continuum limit, this deformation can be regarded as an integrable deformation of the lattice model, which is interesting in its own right.\par

The rest of the paper is organized as follows. In section~\ref{sec:review}, we review the lattice regularization of sine-Gordon model. In section~\ref{sec:continuum}, we discuss the $T\bar{T}$-deformation of the sine-Gordon model in the continuum limit. In section~\ref{sec:dynamic}, we propose the deformation of the lattice model and show that in the continuum limit it indeed leads to the $T\bar{T}$-deformed sine-Gordon model. We also study the ground state energy of the deformed finite lattice model. We conclude in section~\ref{sec:conclude} and discuss future directions.

%%%%%%%%%%%%%%%%%%%%%%%%%%%%%%%%%%%%%%%%%%%%%%%%%%%%%%%%%%%%%%%%%%%%%%%%
\section{Lattice regularization of IQFT}
\label{sec:review}
%%%%%%%%%%%%%%%%%%%%%%%%%%%%%%%%%%%%%%%%%%%%%%%%%%%%%%%%%%%%%%%%%%%%%%%%
In this section, we review the light-cone lattice regularization of sine-Gordon model \cite{Destri:1987ze,Destri:1994bv}. For a more detailed discussion, see for example \cite{Feverati:2000xa}.

\subsection{Light-cone lattice regularization}
\begin{figure}[h!]
\begin{center}
\includegraphics[scale=0.5]{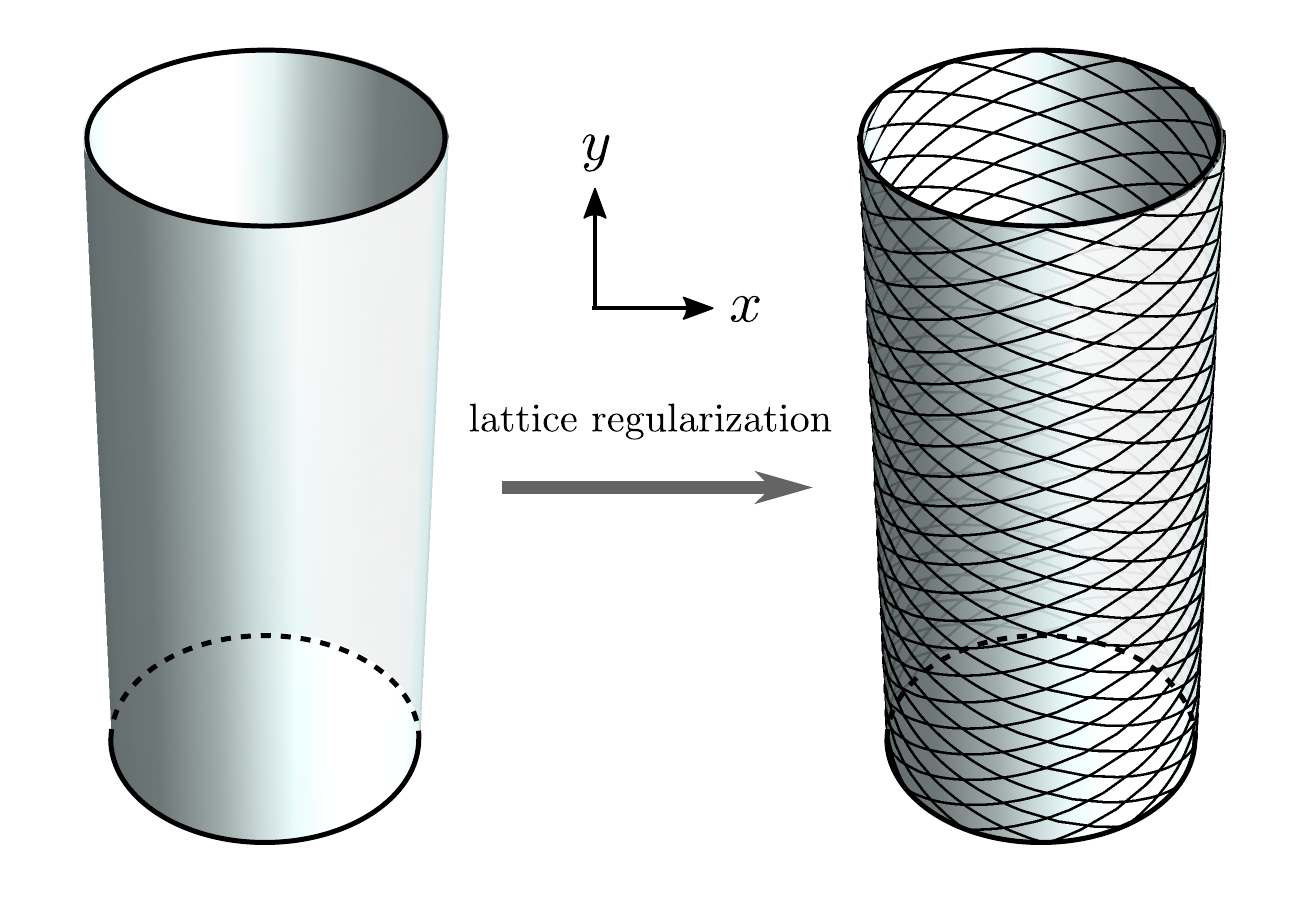}
\caption{Light-cone lattice regularization of spacetime. The spacial direction is compact.}
\label{fig:lattice}
\end{center}
\end{figure}

Let us consider a Minkowski spacetime in two dimensions. We denote the space and time coordinates as $x$ and $y$ and define the light-cone coordinate as $x^{\pm}=y\pm x$. We take the space direction to be compact with length $L$, as is shown in figure~\ref{fig:lattice}.
The lattice spacing in is denoted by $a$. The light-cone lattice is defined by
\begin{align}
\mathcal{M}=\{x_{\pm}={a} n_{\pm}/{\sqrt{2}},n_{\pm}\in\mathbb{Z}\}\,.
\end{align}
We take the spacial direction to have $2N$ sites and $L=N a$.
\paragraph{The Hilbert space} Each spacetime point is associated with 4 links, 2 in the past and 2 in the future. The Hilbert space at each point is denoted by $\mathcal{H}_i$. The generic vector of a basis of $\mathcal{H}_i$ is denoted by
\begin{align}
|\alpha_{2i-1},\alpha_{2i}\rangle\equiv|\alpha_{L_i},\alpha_{R_i}\rangle
\end{align}
where the odd/even numbers refers to right/left moving states. Here each $\alpha_k$ has two degrees of freedom, corresponding to a bare particle which moves towards or away from the spacetime point. The total Hilbert space is
\begin{align}
\mathcal{H}=\otimes_{i=1}^N\mathcal{H}_i.
\end{align}
The basis of $\mathcal{H}$ at fixed time is given by
\begin{align}
|\alpha_1,\alpha_2\rangle\otimes\cdots\otimes|\alpha_{2N-1},\alpha_{2N}\rangle=|\alpha_1,\alpha_2,\cdots,\alpha_{2N}\rangle\in\mathcal{H}.
\end{align}

\paragraph{Dynamics on the lattice} Now we define the dynamics on the lattice. We denote the $S$-matrix associated with $|\alpha_{2i-1},\alpha_{2i}\rangle$ by $S_{2i-1,2i}$. We define the left and right diagonal transfer matrices by $U_L$ and $U_R$ as
\begin{align}
\label{eq:transferLR}
U_L=&\, V S_{12}S_{34}\cdots S_{2N-1,2N},\\\nonumber
U_R=&\, V^{-1} S_{12}S_{34}\cdots S_{2N-1,2N},
\end{align}
where $V$ is the shift operator in the spacial direction by half lattice spacing $a/2$
\begin{align}
V=\rP_{12}\rP_{23}\cdots \rP_{2N-1,2N}.
\end{align}
Here $\rP_{i,i+1}$ is the permutation operator. The transfer matrices \eqref{eq:transferLR} move by one lattice spacing in the left upward and right upward directions. Using these operators, we can define the Hamiltonian $H$ and momentum operator $P$ of the system
\begin{align}
\label{eq:defHP}
e^{-iaH}=U_RU_L,\qquad e^{-iaP}=U_R U_L^{-1}\,.
\end{align}

\subsection{The integrable lattice model}
Now we define an integrable model on the lattice. We take the $S$-matrix at each site to be
\begin{align}
S_{jk}(u)=\rP_{jk}R_{jk}(u)
\end{align}
where $\rP_{jk}$ is the permutation operator and $R_{jk}$ is the $R$-matrix of the XXZ spin chain. More explicitly, the $S$-matrix is given by
\begin{align}
\label{eq:Ssix}
S_{jk}(u)=\left(
         \begin{array}{cccc}
           1 & 0 & 0 & 0 \\
           0 & c(u) & b(u) & 0 \\
           0 & b(u) & c(u) & 0 \\
           0 & 0 & 0 & 1 \\
         \end{array}
       \right)_{jk}
\end{align}
with the Boltzmann weights
\begin{align}
b(u)=\frac{\sinh\kappa u}{\sinh\kappa(i\pi-u)},\qquad c(u)=\frac{i\sin\kappa u}{\sinh\kappa(i\pi-u)}\,.
\end{align}
Here $\kappa$ is a parameter with range $\kappa$ is $0\le \kappa\le 1$. We have chosen the normalization such that for real $u$, the $S$-matrix is unitary
\begin{align}
S_{jk}^{\dagger}(u)S_{jk}(u)=1.
\end{align}
Integrability of the $S$-matrix is guaranteed by the Yang-Baxter equation
\begin{align}
S_{jk}(u)S_{ik}(u+v)S_{ij}(v)=S_{ij}(v)S_{ik}(u+v)S_{jk}(v).
\end{align}
Following the standard algebraic Bethe ansatz, we define the monodromy matrix
\begin{align}
M_a(u|\{\theta_i\})=S_{a1}(u+\theta_1)S_{a2}(u+\theta_2)\cdots S_{aN}(u+\theta_N)
\end{align}
and the transfer matrix
\begin{align}
\label{eq:transferT}
T(u|\{\theta_i\})=\tr_a M_a(u|\{\theta_i\})\,,
\end{align}
where $u$ is called the spectral parameter and $\{\theta_i\}$ are the inhomogeneities. For any choice of $\{\theta_i\}$, one can diagonalize this transfer matrix by Bethe ansatz.

The crucial fact is that for specific choice of the spectral parameter and inhomogeneities, the transfer matrix \eqref{eq:transferT} is related to the shift operators $U_L$ and $U_R$ defined in \eqref{eq:transferLR}. To this end, we take the inhomogeneities to be
\begin{align}
\label{eq:theta}
\theta_i=(-1)^{i+1}\Theta/2,\qquad i=1,2,\cdots,2N.
\end{align}
We denote the corresponding transfer matrix as $T(u|\Theta/2)$. The shift operators are given by
\begin{align}
U_L(\Theta)=T(\Theta/2|\Theta/2),\qquad U_R^{\dagger}(\Theta)=T(-\Theta/2|\Theta/2).
\end{align}
Using Bethe ansatz, we can find the eigenvalues of $T(u|\{\theta_i\})$ with any spectral parameter and inhomogeneities, including the special choice \eqref{eq:theta}. This allows us to diagonalize $U_L$ and $U_R$, which in turn gives the eigenvalues of $H$ and $P$ in \eqref{eq:defHP}.

\subsection{The non-linear integral equation}
Now we discuss how to solve the integrable lattice model. By solving the lattice model, we mean diagonalizing the transfer matrix $T(u|\Theta/2)$. This can be done by Bethe ansatz. Each eigenstate $|\{u_k\}\rangle$ of the transfer matrix is parameterized by $M$ parameters called rapidities
\begin{align}
\{u_k\}=\{u_1,u_2,\cdots,u_M\}.
\end{align}
The rapidities satisfy Bethe ansatz equations (BAE)
\begin{align}
\label{eq:BAE}
\left(\frac{\sinh\kappa(u_j+\Theta+\frac{i\pi}{2})\sinh\kappa(u_j-\Theta+\frac{i\pi}{2})}{\sinh\kappa(u_j+\Theta-\frac{i\pi}{2})\sinh\kappa(u_j-\Theta-\frac{i\pi}{2})}\right)^N
=-\prod_{k=1}^M\frac{\sinh\kappa(u_j-u_k+i\pi)}{\sinh\kappa(u_j-u_k-i\pi)},\quad j=1,\ldots,M.
\end{align}
In terms of the rapidities, the eigenstates of the Hamiltonian and momentum are given by
\begin{align}
e^{ia(E\pm P)/2}=(-1)^M\prod_{j=1}^M\frac{\sinh\kappa(\Theta\pm u_j+i\pi/2)}{\sinh\kappa(\Theta\pm u_j-i\pi/2)}.
\end{align}
In principle, one needs to solve the Bethe equations and find the Bethe roots. In the thermodynamic limit, this is not feasible. One nice method to find the spectrum with $N\gg 1$ is the non-linear integral equation approach. This method plays an important role in establishing relation between the integrable lattice model and the sine-Gordon model. 
\paragraph{The counting function} In order to write down the non-linear integral equation, we define an important quantity called the counting function. For later convenience, let us define the function
\begin{align}
\phi_{\nu}(u)=i\log\left(\frac{\sinh\kappa(i\pi\nu+u)}{\sinh\kappa(i\pi\nu-u)}\right).
\end{align}
The counting function is given by
\begin{align}
\label{eq:ZNdef}
Z_N(u)=N\big[\phi_{1/2}(u+\Theta)+\phi_{1/2}(u-\Theta)\big]-\sum_{k=1}^M\phi_1(u-u_k).
\end{align}
In terms of the counting function, the BAE (\ref{eq:BAE}), or more precisely its logarithmic form becomes
\begin{align}
Z_N(u_j)=2\pi I_j,\qquad I_j\in\mathbb{Z}+\frac{1+\delta}{2}
\end{align}
where $\delta=0,1$ for even and odd $M$ respectively. In order words, if $M$ is even, $I_j$ are half of odd integers; if $M$ is odd, $I_j$ are integers. The $I_j$'s are called mode numbers.

\paragraph{The antiferromagnetic vacuum} In the continuum limit which we will discuss below, different solutions of BAE correspond to different physical states of the sine-Gordon theory. In particular, the ground state of sine-Gordon theory corresponds to the antiferromagnetic vacuum solution of the lattice model. The total spin for the antiferromagnetic vacuum is zero. Therefore we have $M=N$. The mode numbers for the vacuum solution are chosen such that the BAE become
\begin{align}
\label{eq:BAEZN}
Z_N(u_j)=(N-2j+1)\pi,\qquad j=1,2,\cdots,N.
\end{align}
There's a unique solution to this choice of mode number and all the Bethe roots are real numbers and distribute symmetrically with respect to the origin.

Excited states of the sine-Gordon model correspond to other solutions of BAE of the lattice model. For example, if one root is missing in (\ref{eq:BAEZN}), then we have $N-1$ Bethe roots. This is like a `hole' in the Fermi sea and corresponds to the one soliton state of the sine-Gordon model. Similarly, two holes correspond to the two soliton state and so on.

\paragraph{The non-linear integral equation} Using the contour deformation trick, we can rewrite (\ref{eq:ZNdef}) as an integral equation for the counting function. For the detailed derivation, we refer to \cite{Destri:1994bv,Feverati:2000xa}. The result is
\begin{align}
\label{eq:DDVlattice}
Z_N(u)=2N\arctan\left(\frac{\sinh u}{\cosh\Theta}\right)&-i\int_{-\infty}^{\infty}dv\,G(u-v-i\eta)\log\big(1+e^{iZ_N(v+i\eta)}\big)\\\nonumber
&+i\int_{-\infty}^{\infty}dv\,G(u-v+i\eta)\log\big(1+e^{-iZ_N(v-i\eta)}\big)
\end{align}
where the parameter $\eta$ is a small real parameter such that $0<\eta<\pi\kappa/2$. The convolution kernel $G(u)$ is given by
\begin{align}
G(u)=\int_{-\infty}^{\infty}\frac{dk}{4\pi}\frac{\sinh\left[\pi k(\xi-1)/2\right]}{\sinh(\pi k\xi/2)\cosh(\pi k/2)}e^{iku}
\end{align}
The parameter $\xi$ is related to $\kappa$ as
\begin{align}
\kappa=1/(1+\xi).
\end{align}
%We can also write DDV equation in a more compact form
%\begin{align}
%Z_N(u)=2N\arctan\left(\frac{\sinh u}{\cosh\Theta}\right)+2\text{Im}\int_{-\infty}^{\infty}dv\,G(u-v-i\eta)\log\big[1+e^{iZ(v+i\eta)}\big]
%\end{align}
%For later use, we want to point out that this kernel is the thermodynamic Bethe ansatz (TBA) kernel of the sine-Gordon theory and is related to the XXZ TAB kernel by a particle-hole transformation. For any integrable $S$-matrix, the TBA kernel is defined as the logarithmic derivative. We have
%\begin{align}
%G(u-v)=\frac{i}{2\pi}\frac{\partial}{\partial u}\log S_{\text{sG}}(u-v)
%\end{align}
%where $S_{\text{sG}}(u-v)$ is the $S$-matrix of the sine-Gordon theory. The $S$-matrix of the XXZ spin chain is given by
%\begin{align}
%S_{\text{XXZ}}(u-v)=\frac{\sinh\kappa(u-v-i\pi)}{\sinh\kappa(u-v+i\pi)},
%\end{align}
%We define the corresponding TBA kernel as
%\begin{align}
%K(u-v)=\frac{i}{2\pi}\frac{\partial}{\partial u}\log S_{\text{XXZ}}(u-v).
%\end{align}
%Then the two kernels are related to the following particle-hole transformation
%\begin{align}
%G=K\ast (K+1)^{-1}
%\end{align}
%Physically, this relation is a direct result of what we discussed before. Namely, the solitons in the sine-Gordon model can be interpreted as holes on the antiferromagnetic vacuum of the XXZ model. For more detailed derivation, we refer to appendix~\ref{sec:phtrans}.

\paragraph{Energy and momentum}
In terms of the Bethe roots, the energy and momentum of the antiferromagnetic vacuum is given by
\begin{align}
\label{eq:EPinU}
E_N=&\,\frac{1}{a}\sum_{j=1}^N\big[\phi_{1/2}(\Theta-u_j)+\phi_{1/2}(\Theta+u_j)-2\pi\big],\\\nonumber
P_N=&\,\frac{1}{a}\sum_{j=1}^N\big[\phi_{1/2}(\Theta-u_j)-\phi_{1/2}(\Theta+u_j)\big].
\end{align}
Here the choice of branch for the logarithm for $E_N$ is made such that the contribution of each real root is negative definite, so that excitations like holes will have positive energies. Using the contour deformation trick, these can also be written in terms of the counting function
\begin{align}
\label{eq:EP}
E_N=&\,-\frac{1}{a}\,2\text{Im}\int\frac{dv}{2\pi}\left(\frac{1}{\cosh(\Theta-v-i\eta)}-\frac{1}{\cosh(\Theta+v+i\eta)}\right)
\log\big(1+e^{iZ_N(v+i\eta)}\big)\\\nonumber
&\,-\frac{N}{a}(\pi+\kappa\pi),\\\nonumber
P_N=&\,-\frac{1}{a}\,2\text{Im}\int\frac{dv}{2\pi}\left(\frac{1}{\cosh(\Theta-v-i\eta)}+\frac{1}{\cosh(\Theta+v+i\eta)}\right)\log\big(1+e^{iZ_N(v+i\eta)}\big)
\end{align}
Sometimes the constant piece in the energy is also denoted as $E_{\text{bulk}}$
\begin{align}
E_{\text{bulk}}=-\frac{N}{a}\pi(1+\kappa).
\end{align}
In the continuum limit, this term diverges as $N^2$ and is discarded. What we call the Casimir energy in the continuum limit is $E_N-E_{\text{bulk}}$. Equivalently, we can define the shifted energy $\tilde{E}_N$ such that in the continuum limit it directly give the Casimir energy. We have
\begin{align}
\label{eq:EtinU}
\tilde{E}_N=\frac{1}{a}\sum_{j=1}^N\big[\phi_{1/2}(\Theta-u_j)+\phi_{1/2}(\Theta+u_j)+(\kappa-1)\pi\big]
\end{align}
To sum up, to find the energy of the antiferromagnetic vacuum of the lattice model, we first solve the equation \eqref{eq:DDVlattice} to find the counting function $Z_N(u)$ and then compute the energy and momentum by \eqref{eq:EP}.

\subsection{The continuum limit}
We have introduced the light-cone integrable lattice model and discussed its Bethe ansatz solution. To make contact with sine-Gordon theory, we need to take the continuum limit. On the lattice, we have the lattice spacing $a$ which need to be sent to zero. We have the number of sites $N$, which needs to be sent to infinity. We also have the parameter $\Theta$, which plays the role of \emph{cut-off rapidity}. In the continuum limit, we have the length of finite volume $R$, and the renormalized mass scale $m$. It is clear that we have
\begin{align}
R=a N\,.
\end{align}
In the continuum limit we send $a\to 0$ and $N\to\infty$ with $R$ fixed and finite. The cut-off rapidity $\Theta$ is related to the renormalized mass scale $m$. It turns out in the continuum limit we need to send $\Theta\to\infty$ such that
\begin{align}
\label{eq:mtheta}
m\sim \frac{e^{-\Theta}}{a}
\end{align}
is fixed and finite. More precisely, we shall take
\begin{align}
\label{eq:Thetacon}
\Theta=\log\left(\frac{4N}{m R}\right)
\end{align}
and send $N$ to infinity. \par

Plugging (\ref{eq:Thetacon}) into (\ref{eq:DDVlattice}) and taking the limit $N\to\infty$, we obtain
\begin{align}
Z(u)=mR\sinh u+2\text{Im}\int_{-\infty}^{\infty}dv\,G(u-v-i\eta)\log\big[1+e^{iZ(v+i\eta)}\big].
\end{align}
This is exactly the NLIE for the sine-Gordon theory which can be used to determine the finite volume energy spectrum. The NLIE can be solved numerically. Once we find the counting function, the energy and momentum are given by
\begin{align}
E(R)=&\,-2m\,\text{Im}\int_{-\infty}^{\infty}\frac{dv}{2\pi}\sinh(v+i\eta)\log\big[1+e^{iZ(v+i\eta)}\big],\\\nonumber
P(R)=&\,-2m\,\text{Im}\int_{-\infty}^{\infty}\frac{dv}{2\pi}\cosh(v+i\eta)\log\big[1+e^{iZ(v+i\eta)}\big].
\end{align}
The excited states are given by similar equations. We only need to modify the driving terms, or equivalently, deforming the integration contour to encircle certain poles. The equation for excited states can be found for example in \cite{Feverati:1998dt}. For our purpose, which is finding a $T\bar{T}$-like deformation on the lattice, it is sufficient to consider the ground state NLIE. Generalization to excited states is straightforward.

%%%%%%%%%%%%%%%%%%%%%%%%%%%%%%%%%%%%%%%%%%%%%%%%%%%%%%%%%%%%%%%%%%%%%%%%
\section{$T\bar{T}$-deformation in the continuum limit}
\label{sec:continuum}
%%%%%%%%%%%%%%%%%%%%%%%%%%%%%%%%%%%%%%%%%%%%%%%%%%%%%%%%%%%%%%%%%%%%%%%%
In this section, we give a brief review of $T\bar{T}$-deformation. At the Lagrangian level, it is defined as a family of models parametrized by the deformation parameter $t$
\begin{align}
\frac{\partial\mathcal{L}_t}{\partial t}=T\bar{T}_t
\end{align}
where $T\bar{T}$ is a composite operator $\det T_{\mu\nu}$ which can be defined more carefully by point splitting \cite{Zamolodchikov:2004ce}. This deformation is particularly simple for IQFTs, as it amounts to deforming the $S$-matrix by multiplying a CDD factor
\begin{align}
S(u,v)\to S(u,v)S_{\text{CDD}}(u,v).
\end{align}
For the sine-Gordon model, the CDD factor is simply
\begin{align}
\label{eq:SCDD}
S_{\text{CDD}}(u,v)=e^{it m^2\sinh(u-v)}.
\end{align}
Multiplying the $S$-matrix with a CDD factor apparently preserves integrability because the deformed $S$-matrix still satisfies Yang-Baxter equation. In addition, the CDD factor does not introduce new poles on the physical strip and therefore it does not modify the IR physics. Since integrability is preserved, part of the integrability toolkit can be used. In particular, for the sine-Gordon model, it has been shown that the deformed spectrum can be found by the deformed NLIE \cite{Cavaglia:2016oda}. For the ground state, the $T\bar{T}$-deformed NLIE is given by
\begin{align}
\label{eq:TTbarDDV1}
Z(u)=&\,m\sinh(u)\left[R+t E(R,t)\right]+m \cosh(u)\, t P(R,t)\\\nonumber
&\,+2\text{Im}\int_{-\infty}^{\infty}dv\,G(u-v-i\eta)\log\big[1+e^{iZ(v+i\eta)}\big].
\end{align}
where $E(R,t)$ and $P(R,t)$ are the deformed energy and momentum respectively. We can recast the driving term in a more instructive form and write the NLIE as
\begin{align}
\label{eq:TTbarDDV}
Z(u)=&\,mR_t\sinh(u+\varphi_t)+2\text{Im}\int_{-\infty}^{\infty}dv\,G(u-v-i\eta)\log\big[1+e^{iZ(v+i\eta)}\big].
\end{align}
The new parameters $R_t$ and $u_t$ are defined as
\begin{align}
\label{eq:Ru0}
R_t\,\cosh \varphi_t=&\,R+t E(R,t),\\\nonumber
R_t\,\sinh \varphi_t=&\,t P(R,t).
\end{align}
Given that NLIE of the sine-Gordon model can be derived from the NLIE underlying lattice model by taking the continuum limit,  a natural question is:  Is there a deformation of the integrable lattice model  whose continuum limit gives the deformed NLIE (\ref{eq:TTbarDDV}) ?

\section{Integrable deformation on the lattice}
\label{sec:dynamic}
%%%%%%%%%%%%%%%%%%%%%%%%%%%%%%%%%%%%%%%
In this section, we give an affirmative answer to the question we posed at the end the previous section. Since $T\bar{T}$-deformation is an irrelevant deformation, it changes the UV physics. Therefore it is natural to suspect that it is somehow related to the UV cut-off in a certain way. In our lattice model, the UV cut-off is related to the cut-off rapidity $\Theta$ and the lattice spacing $a$. According to \eqref{eq:mtheta}, they are not independent if we assume that the mass $m$ is not modified by the deformation.\par

\subsection{The proposal}
The key idea is to deform the choice of inhomogeneities, which are related to the momenta of the `bare particles'. In the undeformed case, we take the inhomogeneities as in (\ref{eq:theta}). Consider the following choice of the inhomogeneities
\begin{align}
\theta_n(t)=(-1)^{n+1}\big(\Theta+\sigma(t)\big)+\mu(t).
\end{align}
where $\sigma(t)$ and $\mu(t)$ depend on the deformation parameter $t$. Ignoring $\mu(t)$ which is a global shift for all particles, we see that this modification amounts to change the cut-off rapidity from $\Theta$ to $\Theta+\sigma(t)$.
By straightforward computation, we arrive at the following deformed NLIE on the lattice
\begin{align}
Z_N^{(t)}(u)=2N\arctan\left(\frac{\sinh \big(u+\mu(t)\big)}{\cosh\big(\Theta+\sigma(t)\big)}\right)+2\text{Im}\int_{-\infty}^{\infty}dv\,G(u-v-i\eta)\log\big[1+e^{iZ(v+i\eta)}\big]
\end{align}
Taking the continuum limit as before, namely take $\Theta=\log(4N/mR)$ and send $N\to\infty$, we obtain the following deformed NLIE
\begin{align}
Z^{(t)}(u)=m(Re^{-\sigma(t)})\sinh\big(u+\mu(t)\big)+2\text{Im}\int_{-\infty}^{\infty}dv\,G(u-v-i\eta)\log\big[1+e^{iZ(v+i\eta)}\big].
\end{align}
Comparing this equation with the NLIE of the $T\bar{T}$-deformed sine-Gordon theory (\ref{eq:TTbarDDV}), we find that they become the same if we make the following identification
\begin{align}
R_t=R e^{-\sigma(t)},\qquad \varphi_t=\mu(t)\,.
\end{align}
This leads to the following choice of $\sigma(t)$ and $\mu(t)$
\begin{align}
\sigma(t)=\log\left(\frac{R}{R_t}\right),\qquad \mu(t)=\text{arcsinh}\left(\frac{tP}{R_t}\right)
\end{align}
where
\begin{align}
R_t=\sqrt{R^2+2tRE+t^2(E^2-P^2)}
\end{align}
and $E$ and $P$ are the \emph{deformed} energy and momentum. The following comments are in order.
\begin{enumerate}
\item It is obvious that such a deformation is integrable since we did not modify the $\check{R}$-matrix of the lattice model \eqref{eq:Ssix} and it still satisfies the Yang Baxter equation.
\item As we discussed before, keeping the mass scale $m$ fixed, deforming $\Theta$ is equivalent to deforming lattice spacing $a$ in an energy dependent way. Therefore we can interpret the $T\bar{T}$-deformation as putting the integrable theory on a dynamical lattice. The deformed lattice spacing is $a_t\sim a R_t/R$. This is consistent with the dynamical change of coordinate point of view \cite{Dubovsky:2017cnj,Conti:2018tca}.
\item The nature of \emph{non-locality} is clear from this proposal. In order to deform the cut-off rapidity at each spacetime point, we need to know the energy and momentum of the whole system, which are global quantities.
\end{enumerate}
From the third point, the deformations $\sigma(t)$ and $\mu(t)$ depends on energy and momentum. At the same time, the energy and momentum also depend on these deformations. Therefore, we need to calculate these quantities in a self-consistent way. The deformed NLIE and thermodynamic Bethe ansatz like equation in the continuum limit has been investigated in several works \cite{Cavaglia:2016oda,Camilo:2021gro}. In what follows, we show that even at finite $N$, the deformed lattice model can also be solved in a consistent manner.

%\textcolor{red}{YF: It would be nice to do the following things
%\begin{itemize}
%\item Derive a lattice version of the dynamical change of coordinates
%\begin{align}
%\frac{\partial X^{\mu}}{\partial x^{\nu}}=\delta^{\mu}_{\nu}+t\,(\tilde{T}^{(t)})^{\mu}_{\nu}(x)
%\end{align}
%The rough idea is that on the lattice, the coordinates are given by $(x,y)=(n_x a_x,n_y a_y)$ where $n_x, n_y$ are intergers. Now we have the deformed lattice spacing $\tilde{a}_x(t),\tilde{a}_y(t)$ which depend on deformed energy and momentum (similar to the deformed stress tensor). We should be able to write down some equation of the form $\partial{\tilde{a}_\mu(t)}/\partial a_\nu=f_{\mu}^{\phantom{a}\nu}(E,P)$.
%\item Derive a lattice version of the Burgers' equation. In the continuum limit we know that $E(R,t)=E(R+tE,0)$ and write down the Burgers' equation. On the lattice, similarly we have $E(\Theta,t)=E(\Theta+\sigma(t),0)$. This should allow us to write down a similar equation to the Burgers' equation.
%\item Finally it would be great to build a more direct connection between the two proposals. So far we argued that they are consistent, but can we derive one from the other in a more precise way ?
%\end{itemize}
%}

\subsection{Free fermion point}
For simplicity, we consider the free fermion point $\kappa=1/2$ of the DDV equation. At this point, the theory is free. Both analytical and numerical analysis become simpler. 

\paragraph{Undeformed case} For the undeformed theory, the counting function is given by
\begin{align}
Z_N(u)=2N\arctan\left(\frac{\sinh u}{\cosh\Theta}\right)
\end{align}
The Bethe equation for the antiferromagnetic vacuum is
\begin{align}
Z_N(u_j)=(N-2j+1)\pi,\qquad j=1,2,\cdots,N.
\end{align}
The Bethe roots can be found explicitly
\begin{align}
\label{eq:undefBR}
u_j=\text{arcsinh}\left[\cosh\Theta\cot\left(\frac{(2j-1)\pi}{2N}\right)\right],\qquad j=1,2,\ldots,N\,.
\end{align}
%The density of roots for large $N$ is given by
%\begin{align}
%\rho_N(u)=\frac{2N}{\pi}\frac{\cosh u\cosh\Theta}{\cosh(2u)+\cosh(2\Theta)}.
%\end{align}
The momentum and energy can be computed straightforwardly by using \eqref{eq:EPinU} and \eqref{eq:EtinU}. In particular, for $\kappa=1/2$, we have
\begin{align}
\label{eq:energyUndef}
\tilde{E}_N=&\,\frac{1}{a}\sum_{j=1}^N\left[\phi_{1/2}(\Theta-u_j)+\phi_{1/2}(\Theta+u_j)-\frac{\pi}{2}\right]\\\nonumber
=&\,\frac{1}{a}\sum_{j=1}^N\left(2\arctan\left(\frac{\sinh\Theta}{\cosh u_j}\right)-\frac{\pi}{2} \right)
\end{align}
Plugging in the Bethe roots \eqref{eq:undefBR} into \eqref{eq:energyUndef}, we obtain
\begin{align}
\tilde{E}_N
=\frac{1}{a}\sum_{j=1}^N\left(2\arctan\left(\frac{\sinh\Theta}{\sqrt{1+\cosh^2\Theta\cot^2[(2j-1)\pi/(2N)]}}\right)-\frac{\pi}{2}\right)
\end{align}

\paragraph{Deformed Bethe roots} To simplify the analysis, we consider the case where the total momentum of the state $P_N=0$. In fact since we are considering the ground state, the zero momentum condition is automatically satisfied. In this case, the integrable deformation for the lattice model for finite $N$ is simply changing the cut-off rapidity $\Theta$ to
\begin{align}
\label{eq:defTheta}
\Theta\to\Theta_t\equiv \Theta+\sigma(t),\qquad \sigma(t)=-\log\left(\frac{R_t}{R}\right)=-\log\left(\frac{R+t\tilde{E}_N^{(t)}}{R}\right).
\end{align}
Here $\tilde{E}_N^{(t)}$ is the deformed energy
\begin{align}
\label{eq:defFFE}
\tilde{E}_N^{(t)}=\frac{1}{a}\sum_{j=1}^N\left[\phi_{1/2}(\Theta_t-u_j(t))+\phi_{1/2}(\Theta_t+u_j(t))-\frac{\pi}{2}\right]\,.
\end{align}
The deformed Bethe equation takes the same form
\begin{align}
Z_N^{(t)}(u_j)=(N-2j+1)\pi,\qquad j=1,2,\cdots,N.
\end{align}
with
\begin{align}
Z_N^{(t)}(u)=2N\arctan\left(\frac{\sinh u}{\cosh\Theta_t}\right)\,.
\end{align}
The deformed Bethe roots $u_j(t)$ then take the same form as the undeformed case with $\Theta$ replaced by $\Theta_t$
\begin{align}
\label{eq:deformedU}
u_j(t)=\text{arcsinh}\left[\cosh\Theta_t\cot\left(\frac{(2j-1)\pi}{2N}\right)\right],\qquad j=1,2,\ldots,N\,.
\end{align}
Now plugging \eqref{eq:deformedU} back to \eqref{eq:defFFE}, we see that (\ref{eq:defFFE}) can be viewed as an equation for $\tilde{E}_N^{(t)}$. This equation is rather complicated and in general can only be solved numerically.

\begin{figure}[h!]
\begin{center}
\includegraphics[scale=0.5]{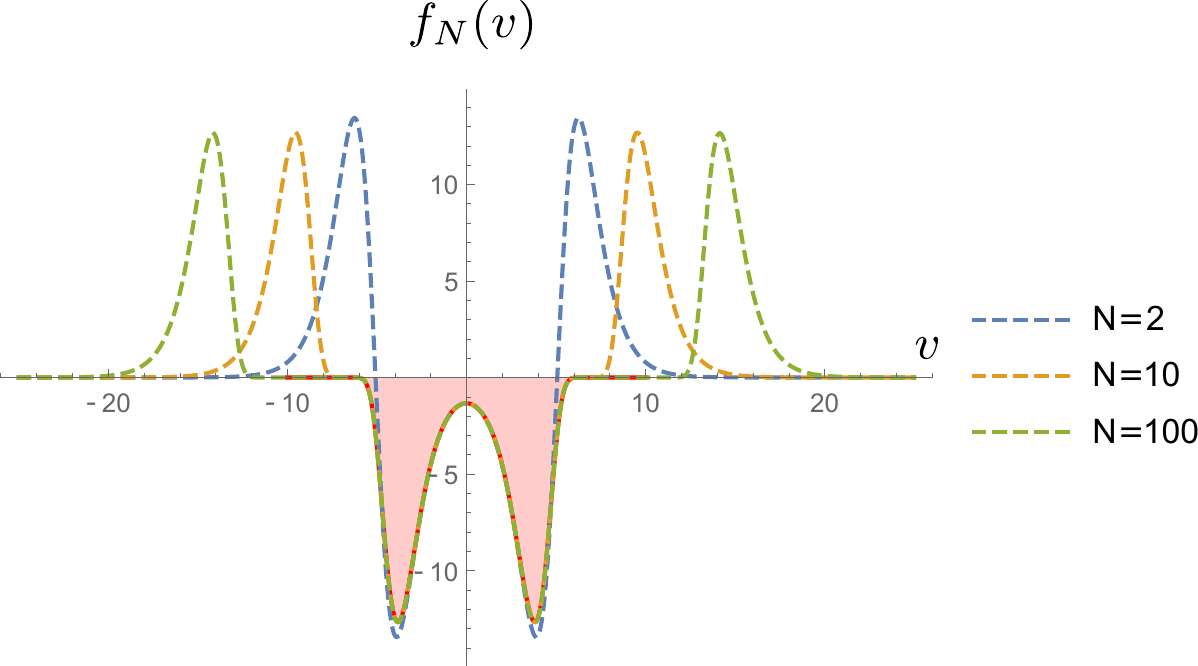}
\caption{Plot of $f_N(v)$ with $m=1,R=0.05,t=0$ at $N=2,10,100$.}
\label{fig:evenN}
\end{center}
\end{figure}

\begin{figure}[h!]
\begin{center}
\includegraphics[scale=0.5]{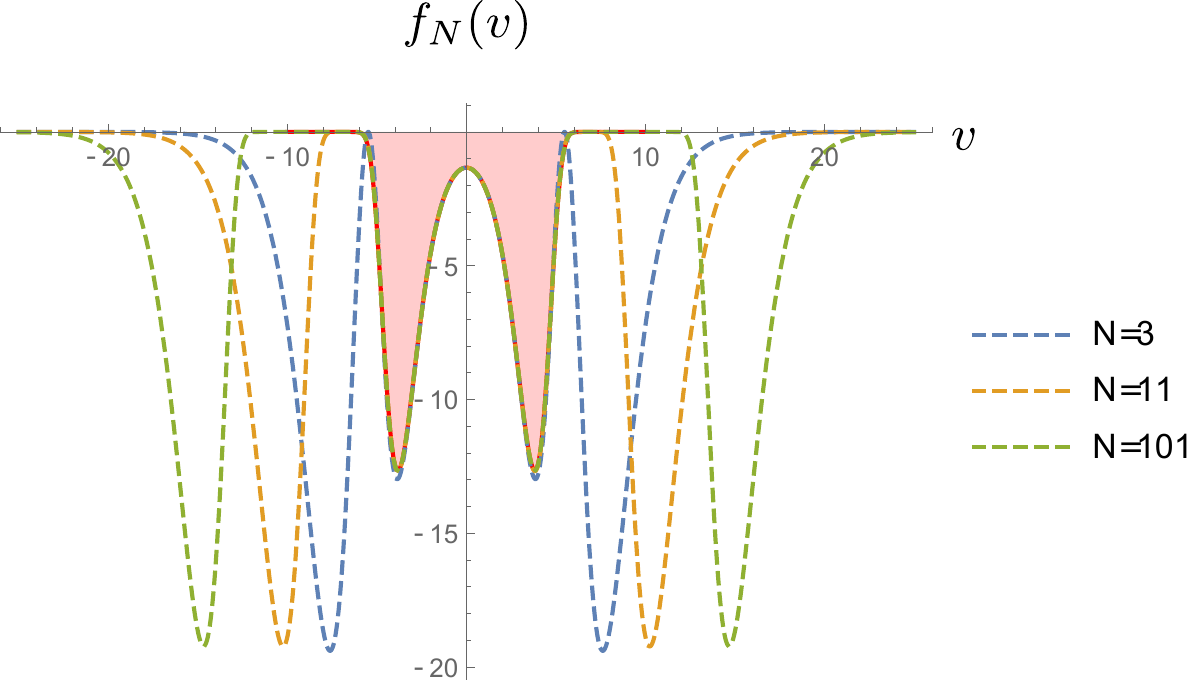}
\caption{Plot of $f_N(v)$ with $m=1,R=0.05,t=0$ at $N=3,11,101$.}
\label{fig:oddN}
\end{center}
\end{figure}

\paragraph{Large $N$ analysis} To see the large $N$ behavior, it is more convenient to use an alternative integral expression for (\ref{eq:defFFE}). Using the contour integration trick, we can write it as
\begin{align}
\label{eq:defintN}
\tilde{E}_N^{(t)}=-\frac{2N}{R}\int_{-\infty}^{\infty}\frac{dv}{2\pi}\frac{4\sinh\Theta_t\cosh v}{\cosh(2\Theta_t)-\cosh(2v)}\log\left[1+e^{-2N \text{arctanh}\left(\frac{\cosh v}{\cosh\Theta_t}\right)} \right]
\end{align}
In the continuum limit, we take $\Theta=\log\frac{4N}{mR}$ with $N\to\infty$. It is straightforward to check that (\ref{eq:defintN}) becomes
\begin{align}
E(R,t)=-2m\int_{-\infty}^{\infty}\frac{dv}{2\pi}\cosh v\log\left[1+e^{-mR_t\cosh v}\right]
\end{align}
which is exactly the expression for the deformed energy. To see how this happens in more detail, let us define the integrand as
\begin{align}
f_N(v)=-\frac{2N}{R}\frac{4\sinh\Theta_t\,\cosh v}{\cosh(2\Theta_t)-\cosh(2v)}\log\left[1+e^{-2N \text{arctanh}\left(\frac{\cosh v}{\cosh\Theta_t}\right)} \right]
\end{align}
with $\Theta=\log\frac{4N}{mR}$. The function $f_N(v)$ depends on the parameters $m,R$ and $t$. For finite $N$, the function $f_N(v)$ are different for even and odd $N$. The plot of $f_N(v)$ for several even values of $N$ is given in figure~\ref{fig:evenN}. The shaded area is the contribution in the continuum limit. There are additional contributions from the upper plane at finite $N$. As $N$ increases, these contributions are pushed towards infinity as is shown in the figure.
Similarly, for odd $N$, the plot of $f_N(v)$ is given in figure~\ref{fig:oddN}. Again the red shaded part is the contribution in the continuum limit. There are extra contributions at finite $N$ on the lower half plane, which is pushed toward infinity as $N$ increase.

%\subsection{Small lattice size}
%In this subsection, we consider the deformed spectrum for small value of $N$. More specifically, we consider $N=2$ and $N=3$, as representatives of even and odd $N$. Our strategy is as follows. We take the undeformed inhomogeneity $\Theta=\log\frac{4N}{mR}$. \textcolor{red}{\textbf{YF}:\emph{Maybe for finite $N$, we don't need to take this form of $\Theta$ since it is only for the purpose of taking proper continuum limit, but can play with different values of $\Theta$.}} For fixed value of $m,R,t$, we can find the deformed energy numerically by solving the integral equation (\ref{eq:defintN}) with $\Theta_t$ given in (\ref{eq:defTheta}).
%
%It is convenient to define the following function $g_E(x)$
%\begin{align}
%g_E(x)=-x-\frac{2N}{R}\int_{-\infty}^{\infty}\frac{dv}{2\pi}&\left\{\frac{4\sinh\Theta\cosh v}{\cosh(2\Theta)-\cosh(2v)}\right.\\\nonumber
%&\left.\times\log\left[1+\exp\left[-2N \text{arctanh}\left(\frac{\cosh v}{\cosh[\Theta-\log|1+t x/R|]}\right)\right]\right]\right\}
%\end{align}
%The deformed energy is given by the zeros of $g_E(x)$. For fixed value of $m,R,t$, we plot the function $g_E(x)$ which will give us some intuition about the deformed spectrum.

\paragraph{Finite $N$ analysis} Now we consider the deformed lattice model for finite $N$. In this case, our perspective is studying the integrable deformation of the lattice model instead of the field theory. Our goal is to describe the method of finding the deformed ground state energy and also study some of its features.\par

To this end, let us consider the simplest case $N=2$, which is already sufficient to exhibit the main features. There are two Bethe roots
\begin{align}
u_1(t)=-u_2(t)=u(t)=\text{arcsinh}(\cosh\Theta_t)
\end{align}
The energy is given by
\begin{align}
\label{eq:consistentE2}
\tilde{E}_2^{(t)}=\frac{1}{a}\left(4\arctan\left(\frac{\sinh\Theta_t}{\sqrt{\cosh^2\Theta_t+1}}\right)-\pi\right)
\end{align}
For the lattice model, we can simply take the lattice spacing $a=1$. The deformation of $\Theta_t$ is given by\footnote{Notice that strictly speaking, the deformation parameter $t$ for the lattice model is slightly different from the one of field theory, it is related to the field theory one by $t_{\text{lattice}}=t_{\text{QFT}}/(aN)$. We take $a=1$ according to our convention.}
\begin{align}
\label{eq:defLattice}
\Theta_t=\Theta+\sigma(t),\qquad e^{-\sigma(t)}=1+t\,\tilde{E}_2^{(t)}\,.
\end{align}
Plugging \eqref{eq:defLattice} back to \eqref{eq:consistentE2}, we obtain an equation for $\tilde{E}_2^{(t)}$. For fixed $\Theta$ and $t$, the resulting equation can be solved numerically, which gives us the deformed energy. Below we present the results for $N=2$ with two different choices of $\Theta$. Note that if we consider the continuum limit, we must take $\Theta\sim\log N$. However, if we simply consider the lattice model, $\Theta$ can be an independent parameter.
\begin{figure}[h!]
\centering
\begin{subfigure}[b]{0.4\textwidth}
\centering
\includegraphics[width=\textwidth]{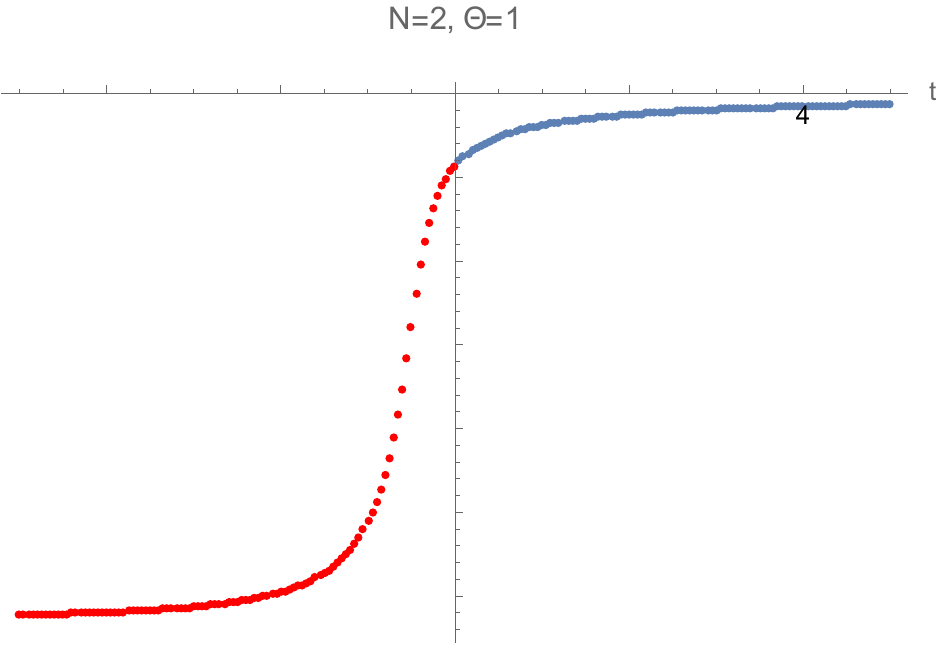}
\caption{Deformed energy with $N=2$, $\Theta=1$.}
\label{fig:N2Theta1}
\end{subfigure}
\hfill
\begin{subfigure}[b]{0.4\textwidth}
\centering
\includegraphics[width=\textwidth]{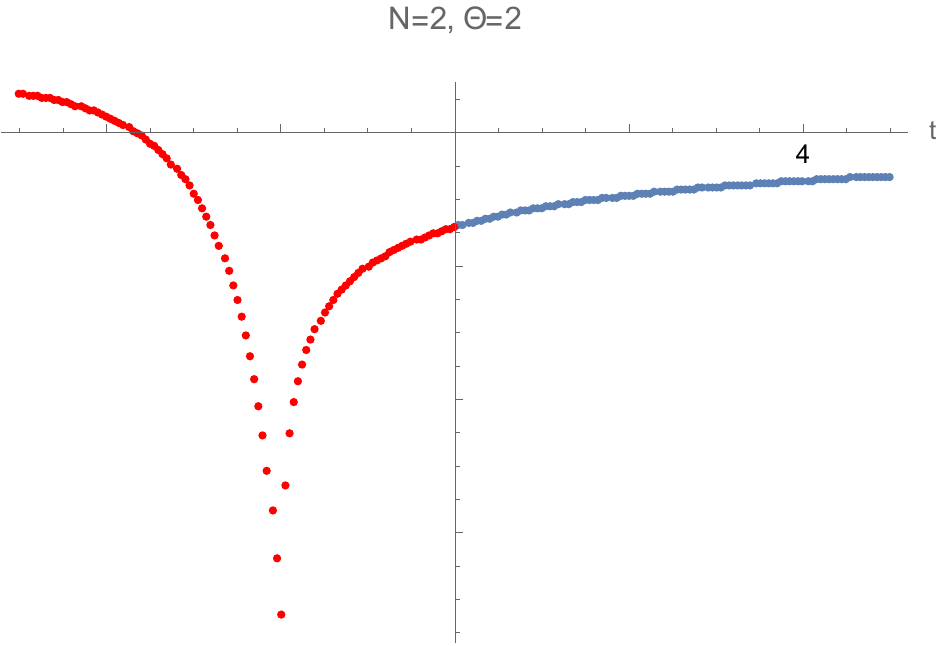}
\caption{Deformed energy with $N=2$, $\Theta=2$.}
\label{fig:N2Theta2}
\end{subfigure}
\caption{Deformed spectrum for $N=2$ with different $\Theta$. The horizontal axis is the deformation parameter $t$ while the vertical axis is the deformed ground state energy $\tilde{E}_2^{(t)}$. The blue and red dots denote the values for positive and negative values of $t$, respectively.}
\end{figure}
We find the deformed energy numerically for various $N$ and $\Theta$. There are two qualitatively different behaviors. For $t>0$, the deformed energy for different $N$ and $\Theta$ have the same behavior which is approaching zero. On the other hand, the behavior for $t<0$ is more interesting. For small $\Theta$, the behavior is given in figure~\ref{fig:N2Theta1} which decreases quickly and then approach a stable value. For sufficiently big $\Theta$, the behavior changes and become the one given in figure~\ref{fig:N2Theta2}, which quickly goes down to a minimum and then increases and approaches some stable value. For larger $N$, we find similar behaviors.

%%%%%%%%%%%%%%%%%%%%%%%%%%%%%%%%%%%%%%%%%%%%%%%%%%%%%%%%%%%%%%%%%%%%%%%%
\section{Conclusions and discussions}
\label{sec:conclude}
%%%%%%%%%%%%%%%%%%%%%%%%%%%%%%%%%%%%%%%%%%%%%%%%%%%%%%%%%%%%%%%%%%%%%%%%
In this paper, we proposed a lattice approach to $T\bar{T}$-deformation of integrable quantum field theory. We proposed an integrability preserving but non-local deformation for the light-cone lattice model whose continuum limit leads to the $T\bar{T}$-deformed sine-Gordon model. The key observation is that, $T\bar{T}$-deformation can be obtained from the lattice model by deforming the cut-off rapidity, or equivalently the lattice spacing in an energy dependent way. This is reminiscent of the dynamical or field dependent coordinate transformation in the field theory.\par

Our proposal at the current stage can be well criticized as being somewhat \emph{ad hoc} because we need to deform the cut-off rapidity in a very specific way. Nevertheless we believe this is a useful first step to gain deeper insights. In fact, this criticism also applies to the dynamical change of coordinate point of view of $T\bar{T}$-deformation of quantum field theory \cite{Conti:2018tca}. However, in quantum field theory the situation is better because there are other equivalent formulations. For example, by coupling the field theory to a JT-like gravity, one can obtain the dynamical change of coordinates in a more natural way by integrating out the gravity degrees of freedom \cite{Dubovsky:2017cnj,Caputa:2020lpa}. Therefore, an important question is whether we have similar formulations on the lattice, namely can we reformulate our proposal as making the lattice dynamical by coupling it to certain lattice gravity ? This idea is similar to putting integrable lattice models on a random lattice, which in the continuum limit results in coupling the corresponding field theory to Liouville gravity, see for example \cite{Kostov:1999qx}.\par

Another important point is that, a given quantum field theory can have different lattice regularizations. For example, one can implement the standard lattice QFT by discretizing fundamental fields like in lattice QCD (see for example the books \cite{Creutz:1983njd,Smit:2002ug}). This is the standard method for putting QFTs on a lattice, but it breaks integrability. In order to have more analytic control over the discretized theory, we would like to preserve integrability. Even under this strict requirement, the discretization is not unique. An alternative discretization of sine-Gordon model was proposed in \cite{Luther:1976mt,Luscher:1976pb}, which relates the discretized sine-Gordon theory to the XYZ spin chain. Given that $T\bar{T}$-deformation is quite universal for 2d quantum field theories, it is a natural question to ask whether we can obtain the $T\bar{T}$-deformed QFT by performing other deformations on different lattice regularizations. There should be some universal features for the deformations of all these regularizations. From the current work, we suspect that the dynamical lattice space picture might play a role in other lattice regularizations as well.\par

Finally, it would be interesting to use our proposal to compute deformed correlation functions. In the current context, some expectation values of local operators and current operators have been computed \cite{Hegedus:2017zkz,Hegedus:2019zks}. It is therefore interesting to compute the corresponding deformed expectation value using our prescription. The results can be compared with other approaches as cross checks.

%%%%%%%%%%%%%%%%%%%%%%%%%%%%%%%%%%%%%%%%%%%%%%%%%%%%%%%%%%%%%%%%
\section*{Acknowledgement}
%%%%%%%%%%%%%%%%%%%%%%%%%%%%%%%%%%%%%%%%%%%%%%%%%%%%%%%%%%%%%%%%
I would like to thank John Cardy, Shota Komatsu and Roberto Tateo for helpful discussions. This work is partly supported by Startup Funding no. 3207022217A1 of Southeast University.

%\bibliographystyle{JHEP}
%\bibliography{yunfeng}

\providecommand{\href}[2]{#2}\begingroup\raggedright\endgroup

\end{document}